\documentclass[12pt]{article}

\pdfoutput=1

\usepackage[affil-it]{authblk}
\usepackage{graphicx}
\usepackage{amsmath,amssymb}
\usepackage{caption}
\usepackage[margin=1cm]{geometry}
\usepackage{color}
\usepackage[super,comma]{natbib}
\begin{document}

\title{Solution to the one-dimensional Rayleigh-Plesset equation by the Differential Transform method}

\author{Aneet Dharmavaram Narendranath%
  \thanks{Electronic address: \texttt{dnaneet@mtu.edu}; Corresponding author}}
\affil{Mechanical Engineering-Engineering Mechanics,\\ Michigan Technological University,\\ Houghton, MI 49931, USA}

%
\date{}

\maketitle

\section{Abstract}

The differential transform method is used to provide a solution to the Rayleigh-Plesset equation in one dimension.  

\section{Background}

The differential transform method (DTM) is a relatively new technique that may be used to find a series solution to differential equations (both linear and nonlinear) through an iterative process.  The reader is directed to accounts by
\citet{Hassan2008a,Momani2008a,Mukherjee2010a} for a detailed analysis of the method.  The Rayleigh-Plesset equation, that describes bubble dynamics, has only recently been solve analytically by \citet{Kudryashov2015a}.  This brief manuscript is an initial effort in applying the DTM to provide a series solution to the one-dimensional Rayleigh-Plesset equation (RPE).

\section{The Rayleigh-Plesset equation in one-dimension}
\begin{figure}
\centering
\includegraphics[width=0.7\textwidth]{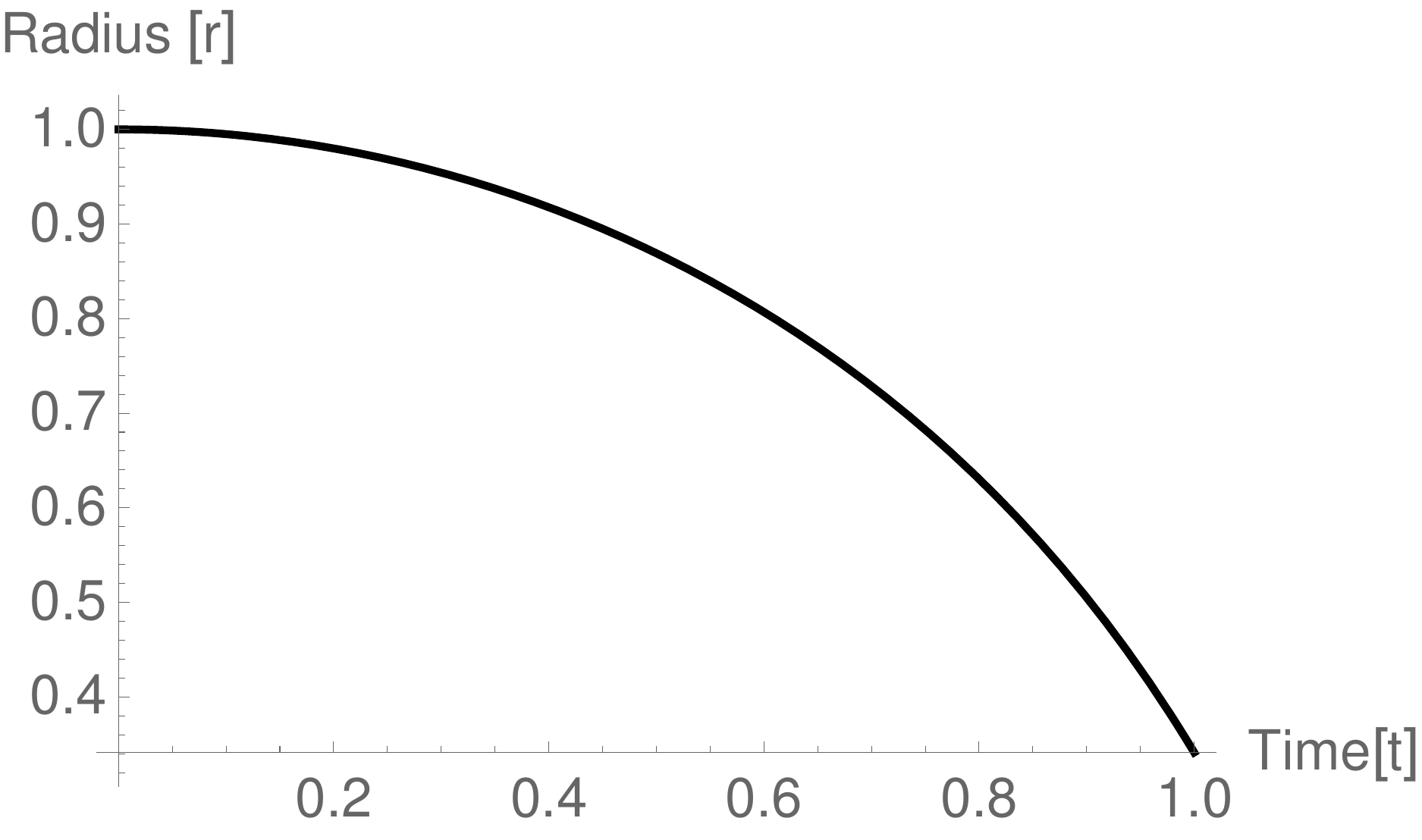}
\caption{Bubble dynamics from numerical solution of the RPE.}
\label{f:1}
\end{figure}

The form of the Rayleigh-Plesset equation solved in this article is given in equation \ref{e:1}
\begin{equation}\label{e:1}
\left(\frac{\mathrm{d}r}{\mathrm{d}t}\right)^2 + \frac{\mathrm{d}^2r}{\mathrm{d}t^2}r + r = 0
\end{equation}
This nonlinear differential equation is solved as an initial value problem with the initial conditions $r(0)=1, r'(0)=0$ where the prime notation is for differentiation.  The numerical solution is obtained using the LSODA method in Wolfram Mathematica's \textbf{NDSolve} function.  This function may be used for heavily stiff nonlinear partial differential equations to provide highly accurate results \citet{Narendranath2014a}.  The bubble radius evolution is depicted in figure \ref{f:1}.

The differential transform method when applied to the RPE as per the rules/operation properties provided in \citet{Hassan2008a} yields equation \ref{e:2}.
\small
\begin{equation}\label{e:2}
\sum\limits_{l=0}^k \left[(l+1)F(l+1)(k-l+1)F(k-l+1) + F(l)(k-l+1)(k-l+2)F(k-l+2)\right]+ F(k)=0
\end{equation}

The first two values of the differential transform, viz., $F(0)$ and $F(1)$ are available from the initial conditions as$F(0)=1, F(1)=0$.  Iterative evaluation of the differential transform in equation \ref{e:2} reveals the following values of $F(k)$ (table \ref{t:1}).

\begin{table}[h]
\centering
\begin{tabular}{ll}
$k$ & $F(k)$\\
\hline
F(0) & 1\\
F(1) & 0\\
F(2) & -0.5\\
F(3) & 0\\
F(4) & -0.0833333\\
F(5) & 0\\
F(6) & -0.0388889\\
F(7) & 0\\
F(8) & -0.0222222\\
F(9) & 0\\
F(10) & -0.0141049
\end{tabular}
\caption{Iterative evaluation of differential transform values}
\label{t:1}
\end{table}
Expressing $r(t) = \sum t^l F(l)$, we have a power series solution to the RPE (equation \ref{e:3}). 
\begin{equation}\label{e:3}
r(t) = 1-0.5t^2-0.08333t^4-0.03888t^6-0.02222t^8-0.01410t^10
\end{equation}

\begin{figure}[h]
\centering
\includegraphics[width=0.9\textwidth]{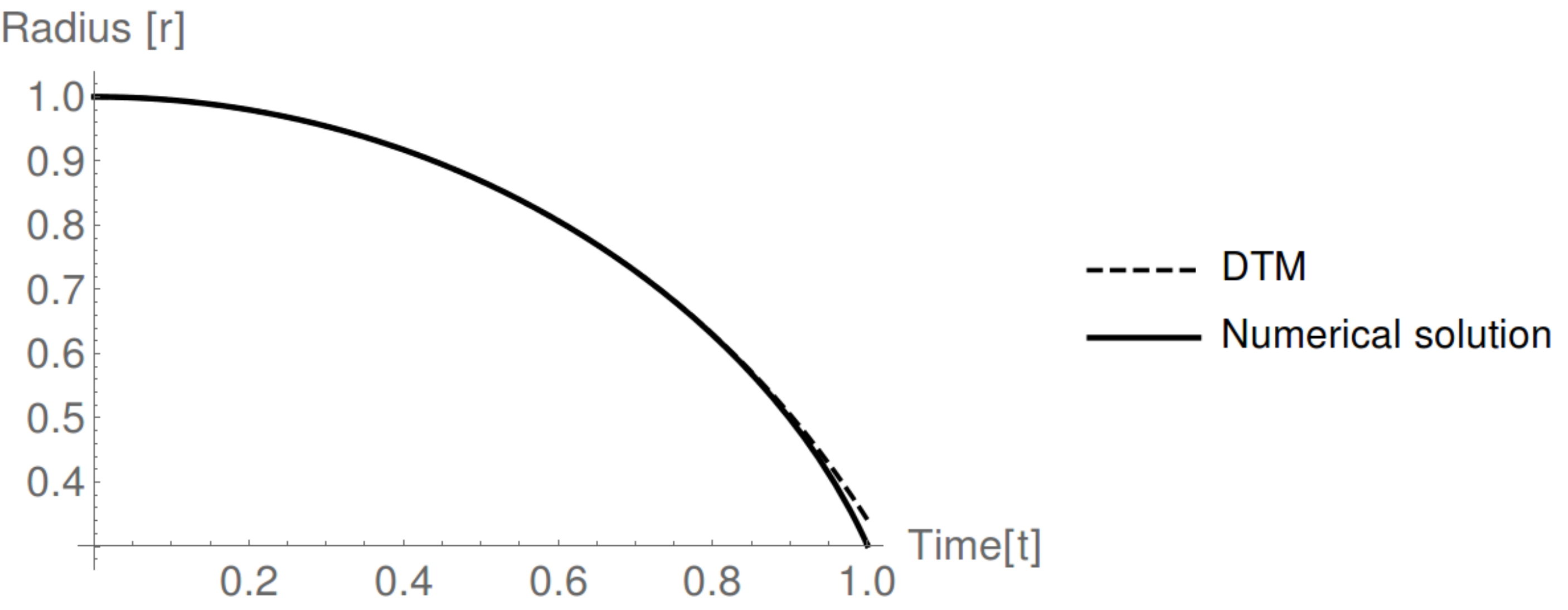}
\caption{Bubble dynamics as captured by the numerical solution of the RPE and by the DTM}
\label{f:2}
\end{figure}

Comparing this power series solutions obtained using the DTM with the numerical solution (table \ref{t:2} and figure \ref{f:2}), we find that for a large duration of the bubble dynamics, the numerical solution and the power series approximation from the DTM have a significantly small difference (measured through percentage error of DTM relative to the numerical approximation).  At the juncture when the bubble collapses ($r\rightarrow 0$), the error is high.  At the time of writing of this article, the value of the iterative variable, $k$ is limited to 10. A larger value would result in generally closer results.

\begin{table}[h]
\centering
\begin{tabular}{cccc}
Time, $t$ & Numerical solution($num$) & DTM($dtm$) & Percentage error $(num-dtm)*100/num$\\
\hline
 0. & 1. & 1. & 0. \\
 0.1 & 0.994991 & 0.994992 & -0.0000151721 \\
 0.2 & 0.979864 & 0.979864 & -0.0000160194 \\
 0.3 & 0.954295 & 0.954295 & -0.0000179592 \\
 0.4 & 0.917691 & 0.917691 & -0.0000394072 \\
 0.5 & 0.86908 & 0.869083 & -0.000350249 \\
 0.6 & 0.806899 & 0.806927 & -0.00350168 \\
 0.7 & 0.728532 & 0.728737 & -0.028076 \\
 0.8 & 0.629204 & 0.630429 & -0.194832 \\
 0.9 & 0.498499 & 0.505174 & -1.33893 \\
 1. & 0.302054 & 0.341451 & -13.0431 \\
\end{tabular}
\caption{Comparison of DTM with the numerical solution}
\label{t:2}
\end{table}

\section{Conclusions}
This short article was an initial foray into applying the Differential transform method (DTM) to find an approximate solution to one form of the Rayleigh-Plesset equation that describes time dynamics of bubbles. The DTM allows for a close match with the numerical solution. Future work will look to use this DTM for more complex problems of two dimensional hydrodynamics.
\bibliographystyle{unsrtnat}
\bibliography{ref}

\end{document}